\documentclass[11pt,twoside,a4paper,cmspaper,final,collab]{cms-tdr}
\def\svnVersion{03e43ec-D}\def\svnDate{2026/07/02}\def\cmsCernNoTag{CERN-EP-2026-155}\def\cmsCernDate{\today}\def\cmsMessage{Submitted to Physics Letters B}
\begin{document}\cmsNoteHeader{HIN-24-015}

\newlength\cmsFigWidth
\ifthenelse{\boolean{cms@external}}{\setlength\cmsFigWidth{0.49\textwidth}}{\setlength\cmsFigWidth{0.6\textwidth}}
\newlength\cmsFigWidthHalf
\ifthenelse{\boolean{cms@external}}{\setlength\cmsFigWidthHalf{0.45\textwidth}}{\setlength\cmsFigWidthHalf{0.5\textwidth}}
\providecommand{\cmsTable}[1]{\resizebox{0.5\textwidth}{!}{#1}}
\newcommand{\PbPb}{\ensuremath{\text{PbPb}}\xspace}
\newcommand{\Ddecay}{\ensuremath{\PDz\to\PKm\Pgpp}\xspace}
\newcommand{\Dbardecay}{\ensuremath{\PADz\to\PKp\Pgpm}\xspace}
\newcommand{\absy}{\ensuremath{\abs{y}}\xspace}
\newcommand{\abseta}{\ensuremath{\abs{\eta}}\xspace}
\newcommand{\pPDz}{\ensuremath{p(\PDz)}\xspace}
\newcommand{\vPDz}{\ensuremath{v_2(\PDz)}\xspace}
\newcommand{\QPOI}{\ensuremath{Q_{2,\text{POI}}}\xspace}

\cmsNoteHeader{HIN-24-015}
\title{Exploring the origin of \texorpdfstring{\PDz}{D0} meson elliptic flow in \texorpdfstring{\PbPb}{PbPb} collisions at \texorpdfstring{$\sqrtsNN=5.02\TeV$}{sqrt(sNN) = 5.02 TeV} using event shape engineering}

\date{\today}

\abstract{The influence of the initial-state geometry on the elliptic flow of prompt \PDz mesons in high-energy heavy ion collisions is explored. A lead-lead (\PbPb) data sample at a center-of-mass energy per nucleon pair of 5.02\TeV and with an integrated luminosity of 0.607\nbinv was collected in 2018 with the CMS detector at the CERN LHC. Based on these data, an event-shape engineering technique is used to isolate collisions with similar geometrical properties. An asymmetry parameter, $q_2$, is first determined from the distribution of transverse energy in the forward region of the detector. This parameter is shown to be linearly correlated with the elliptic anisotropy of inclusive charged particles measured near mid-rapidity, as characterized by the second-order Fourier coefficients, $v_2$. Taking the charged particle $v_2$ coefficients as proxies for the initial-state eccentricity, the correlation of \PDz meson $v_2$ values near mid-rapidity with those of inclusive charged particles is then studied. For \PDz mesons with transverse momenta in the range of 2--30\GeV and for different degrees of collision overlap, the $v_2$ values are found strongly correlated with those for charged particles when selected based on similar $q_2$ intervals. This correlation suggests that the initial-state geometry substantially impacts the development of charm-hadron flow in heavy ion collisions. A model calculation that explores the thermalization and collective motion of \PDz mesons and charged particles is compared to the experimental results.}

\hypersetup{
pdfauthor={CMS Collaboration},%
pdftitle={Exploring the origin of D0 meson elliptic flow in PbPb collisions at sqrt(s[NN]) = 5.02 TeV using event shape engineering},%
pdfsubject={CMS},%
pdfkeywords={CMS, D0, heavy flavor flow, event shape engineering, prompt}}

\maketitle 

\section{Introduction}
Relativistic heavy ion collisions can create a quark-gluon plasma (QGP), a state of matter composed of deconfined quarks and gluons~\cite{QGP}. The QGP has a shear viscosity to entropy density ratio that is close to that of an ideal fluid~\cite{QGP_1} and its evolution can be described by relativistic hydrodynamics. In a near-ideal fluid, the initial spatial geometry of the QGP can be effectively transformed into momentum anisotropies in the final state, leading to collective patterns in azimuthal distributions of emitted particles~\cite{theo_5, flow3}. This azimuthal asymmetry, which has been observed in many heavy ion measurements at both the BNL RHIC and the CERN LHC, can be quantified with Fourier coefficients, $v_n = \langle\cos[n(\phi-\Psi_n)]\rangle$, where $\phi$ represents the azimuthal angle of a particle, $\Psi_n$ denotes the azimuthal angle of the symmetry plane for the $n$th order harmonic~\cite{flow1,flow2}, and the $\langle\rangle$ is the average taken over all particles of interest in all events. The $v_2$ coefficient, commonly referred to as elliptic flow, is expected to exhibit a linear relationship with the corresponding initial-state eccentricity of the QGP for low transverse momentum (\pt) particles~\cite{theo_8, pearson, theo_10, PhysRevC.93.014909}. While $v_2$ values for inclusive charged particles and for identified particles containing light quarks have been extensively studied, detailed investigations of $v_2$ values for particles containing heavy quarks, such as \PDz mesons, offer a novel and important handle to study the properties of the QGP medium. The large mass of heavy quarks renders them sensitive to the early stages of the collision and their subsequent interaction with the medium, providing unique insights beyond those obtainable from light-quark measurements.

Positive \PDz meson $v_2$ values that suggest hydrodynamic flow have been observed at both RHIC~\cite{RHIC1,RHIC2,RHIC3} and LHC~\cite{LHC1,LHC2,LHC3,method2,method1,CMS:2024krd,ALICE_ESE2,ALICE_ESE1} for different collisional centralities. Centrality is defined by the fraction of the total inelastic hadronic cross section, with the 0--10\% centrality representing the 10\% most central collisions and, thus, the largest nuclear overlap. Measurements of \pt-differential $v_2$ values of prompt \PDz mesons in lead-lead (\PbPb) collisions from the CMS experiment demonstrate a clear centrality dependence, which already suggests the influence of the initial-state geometry on the final-state momentum anisotropy of these particles~\cite{method1}. A recent CMS measurement of prompt \PDpms meson anisotropic flow reaches a similar conclusion in the $4<\pt<10\GeV$ region~\cite{2602.14221}. Recent ALICE measurements of the elliptic flow of prompt \PDz, \PDp, \PDps, and \PcgLp $v_2$ in \PbPb collisions provide complementary evidence for the partonic origin of charm-hadron $v_2$ values~\cite{2603.18966}.

The average initial eccentricity of the overlap region in heavy ion collisions is related to the measured collision centrality. However, event-by-event fluctuations in the initial-state shape within a given centrality class lead to a variation in the final-state momentum anisotropy. To study the event-by-event geometry fluctuations at a fixed centrality, an event-shape engineering technique has been developed to select events according to final-state observables that are correlated with the initial-state eccentricity. A second-order flow vector $Q_2$ is first defined as
\begin{equation}\label{eq:Q2}
 Q_2=\sum_{j=1}^M w_j \re^{i 2 \phi_j},
\end{equation}
where the sum is over the multiplicity ($M$) of charged particles at azimuths $\phi_j$. The weights $w_j$ are applied to optimize the resolution of the flow vectors. A related $q_2$ value, which has a reduced multiplicity dependence, is defined as
\begin{equation}\label{eq:q2_reduced}
    q_2=\frac{\abs{\sum_{j=1}^M w_j \re^{i 2 \phi_j}}}{\sum_{j=1}^M w_j }.
\end{equation}
The $q_2$ value reflects the elliptic azimuthal asymmetry of particle emission and can be determined in different pseudorapidity ($\eta$) ranges of the detector, either by using single particle tracks (with $w_j = p_{\text{T,}j}$) or by analyzing the transverse energy ($\ET = E/\cosh(\eta)$) deposits in an azimuthally segmented calorimeter detector (taking $w_j = E_{\text{T},j}$). For a given range of collision centrality, the corresponding $q_2$ distribution can be used to better control the elliptic anisotropy of particle emission. The details of the $q_2$ values used in this analysis are provided in Section~\ref{sec:ESE}. Using such a technique, a connection between the \PDz meson $v_2$ values and the bulk flow in \PbPb collisions at $\sqrtsNN = 5.02\TeV$ was previously found by the ALICE Collaboration based on a $q_2$ analysis~\cite{ALICE_ESE1}.

This Letter presents a measurement of the prompt \PDz meson $v_2$ values determined at mid-rapidity ($\absy<1$) as a function of $q_2$ values measured at forward $\eta$ in \PbPb collisions at a center-of-mass energy per nucleon pair of $\sqrtsNN = 5.02\TeV$. The data were collected in 2018 using the CMS detector at the CERN LHC and correspond to an integrated luminosity of 0.607\nbinv. By classifying events based on $q_2$ values, a systematic variation of the initial-state eccentricity is achieved, enabling the study of initial-state geometry influence on the \PDz meson $v_2$ values. The \pt dependent \PDz meson $v_2$ values, with $2<\pt<30\GeV$, are measured in five centrality classes (0--10, 10--20, 20--30, 30--40, and 40--50\%), each of which is further divided into ten $q_2$-based event-shape classes. The inclusive charged particle $v_2$ values measured for $1<\pt<3\GeV$ are used as an experimental proxy for the event-by-event eccentricity, motivated by the approximately linear hydrodynamic response of low-\pt charged particles to the initial geometry. These inclusive measurements have the same ranges of rapidity, centrality, and $q_2$ as the \PDz mesons. The correlation between \PDz meson $v_2$ values and the low-\pt charged particle $v_2$ values in the ten event-shape classes for each centrality range is studied and compared to a hydrodynamic model. Pearson correlation coefficients~\cite{pearson} are calculated to quantify the linearity of the \PDz meson \vs charged particle $v_2$ correlations. Tabulated results are provided in the HEPData record for this analysis~\cite{hepdata}.

\section{The CMS detector}
The central feature of the CMS apparatus is a superconducting solenoid of 6\unit{m} internal diameter, providing a magnetic field of 3.8\unit{T}. Within the solenoid volume, there is a silicon pixel and strip tracker that measures the trajectories of charged particles in the range $\abseta<2.5$, a lead-tungstate crystal electromagnetic calorimeter, and two forward rapidity hadron calorimeters. For charged particles with $1<\pt<10\GeV$ and $\abseta<1.4$, the typical track resolution is 1.5\% in \pt and 25--90 (45--150)\mum in the transverse (longitudinal) impact parameter~\cite{ppreconstruct}. The two hadron forward (HF) calorimeters use steel as an absorber and quartz fibers as the sensitive material and are segmented to form $0.175{\times}0.175\unit{radians}$ ($\Delta\eta{\times}\Delta\phi$) towers. These calorimeters are located 11.2\unit{m} away from the interaction region, one on either side, and together they extend the calorimeter coverage to $3.0<\abseta<5.2$. In this analysis, the HF information is used for performing the offline event selection, for determining event centrality, with higher energy deposits corresponding to more central events, and for calculating the $q_2$ values. A detailed description of the CMS experiment can be found in Refs.~\cite{CMS:detector,CMS:2024aqx}.

Events of interest are selected using a two-tiered trigger system. The first level, composed of custom hardware processors, uses information from the calorimeters and muon detectors to select events at a rate of around 100\unit{kHz} within a fixed latency of 4\mus~\cite{CMS:2020cmk}. The second level, known as the high-level trigger, consists of a farm of processors running a version of the full event reconstruction software optimized for fast processing, and reduces the event rate to a few kHz before data storage~\cite{CMS:2016ngn,CMS:2024aqx}. 

\section{Event selection and simulated event samples}
The data sample consists of minimum bias (MB) \PbPb events corresponding to an integrated luminosity of 0.607\nbinv~\cite{Lumi2}. The MB selection requires signals above readout thresholds between 6--12\GeV in both of the HF calorimeters~\cite{CMS:2016ngn}. Events are required to pass selection criteria designed to reject background events such as beam-gas interactions~\cite{PbPbreconstruct}. The final results include only events with at least one reconstructed primary vertex (PV) associated with two or more tracks within 15\unit{cm} from the nominal interaction point along the beam direction, and at least two towers with energy larger than 4\GeV in each of the HF detectors. Finally, in order to suppress the contamination from events with multiple ion-ion collisions, an event is rejected if the cluster shapes in the pixel part of the tracker are not compatible with those expected in a single \PbPb collision~\cite{CMS:2024krd}. The total event selection efficiency is $(97\pm2)\%$.

Monte Carlo (MC) generators are employed to investigate the kinematics of \PDz mesons in \PbPb collisions. The \PDz mesons are produced using $\PYTHIA 8.212$~\cite{PYTHIA8} with the CP5 tune~\cite{Eventune}, and their decays are modeled using \EVTGEN 1.3.0~\cite{jetquench}. Samples are generated for both prompt and nonprompt \PDz mesons from \PQb-hadron decays. The decay products of the \PDz mesons are embedded into MB events generated via $\HYDJET$ 1.9~\cite{jetquench}, which is tuned to reproduce the observed charged particle multiplicity and spectra. The CMS detector response to these combined events are simulated using the \GEANTfour package~\cite{GEANT4}.

\section{Signal reconstruction and extraction}\label{sec:sre}
The \PDz mesons are reconstructed using the decay channels \Ddecay and \Dbardecay; in this Letter, these two channels are combined and collectively referred to as \PDz. The tracks used for the reconstruction have $\pt>1\GeV$ and $\abseta<1.2$. Only tracks passing high-purity criteria~\cite{ppreconstruct} are used. The relative \pt resolution of the reconstructed \PDz candidate is less than 10\%. Other requirements for the tracks include having a minimum of 11 hits along the track trajectory in the tracker and having the track fit $\chi^2$ divided by the total number of degrees of freedom of the fit, and also divided by the total number of tracker layers having a hit along the track path, be less than 0.18. The reconstruction is performed by combining pairs of oppositely charged particle tracks with invariant mass $m_\text{inv}$ within ${\pm}130\MeV$ of the world-average \PDz meson mass of 1865\MeV~\cite{pdg}. Since the CMS detector does not provide particle identification information for high-\pt charged particle tracks, both possible particle assignments (pion and kaon) are considered for each pair of selected candidate decay tracks. 

\begin{figure}[!htb]
    \centering
    \includegraphics[width=0.5\textwidth]{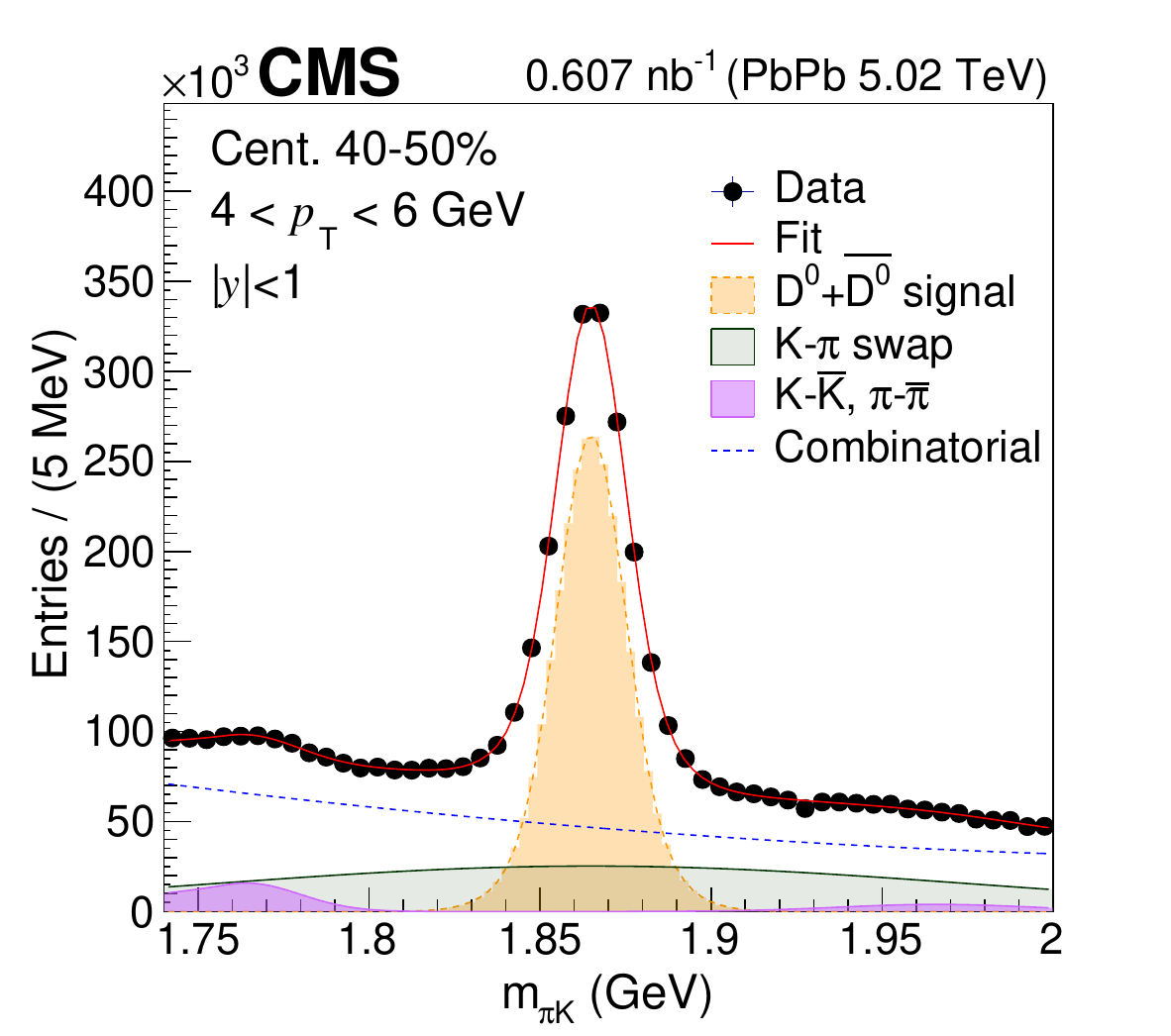} 
    \caption{An example of the fit to the invariant mass spectrum in the centrality class 40--50\% and for the \PDz \pt range 4--6\GeV.}
    \label{fig:massfit}    
\end{figure}

Following the reconstruction of \PDz candidates, a boosted decision tree (BDT) algorithm from the TMVA package~\cite{BDT} is used to optimize the significance of signal, which is given as the ratio of the signal to the signal uncertainty, obtained from fitting the data using the function described below and illustrated in Fig.~\ref{fig:massfit}. For the training, the prompt reconstructed \PDz mesons from MC simulations that match the \PDz particle at the generator level are used as signal candidates, and the side bands from data within the mass range of three to six standard deviations of the MC signal distribution away from the mass peak are used to simulate the background. Training is performed separately for each \pt range and centrality class. The variables used to discriminate the signal from the background are the $\chi^2$ probability for the \PDz vertex, the significance of the distance between the PV and secondary vertices, and the angle between the \PDz momentum and the vector pointing from the PV to the secondary vertex (pointing angle). The variables related to the decay products of the \PDz meson used for the training are the significance of the distances of the closest approach (DCA) to the primary vertex, both along and perpendicular to the beam direction, and the \pt of the decay tracks. Overtraining checks are done for all analysis bins by comparing the BDT distributions from training with those from \PDz meson test samples. The mass spectra are corrected for \PDz acceptance ($A$) and efficiency ($\epsilon$) effects by weighting each \PDz candidate by a factor given by the inverse of the product of acceptance and efficiency. The product is calculated by taking the ratio of the number of reconstructed \PDz meson passing the selection criteria to the total number of generated \PDz mesons in the same kinematic range. To account for a possible discrepancy between the \pt spectra from the MC and data, the MC \pt spectra is weighted with a correction factor so that the MC spectra resemble the one from data. The acceptance and efficiency factor is calculated for each centrality and in finer \pt bins compared to the analysis bins.

The signal extraction is performed by fitting the mass distribution using four components: a second-order polynomial for the combinatorial background, the sum of two Gaussian functions with a common mean but different widths for the \PDz signal to account for the variation of track momentum resolution with $y$, one additional Gaussian function to describe the invariant mass shape due to incorrect mass assignments when swapping the \Pgpp(\Pgpm) and \PKm(\PKp) daughters, and two Crytsal Ball functions~\cite{CrystalBallRef} to describe the decays $\PDz\to\PKm\PKp$ and $\PDz\to\Pgpp\Pgpm$. The widths and the relative areas of the Gaussian functions are obtained from the fit to the simulated signal. To account for the difference in the widths of the Gaussian functions in the simulation and data, each width of the Gaussian functions is multiplied by an additional common scaling factor, obtained by fitting the mass distribution. The swap component and the Crytsal Ball functions do not contribute to the signal yield. An example of the signal extraction is shown in Fig.~\ref{fig:massfit}.

\section{Analysis technique}\label{sec:an_tech}
The scalar product (SP) framework~\cite{scalar1,scalar2} is used to determine the $v_2$ values for all particles of interest, whether \PDz mesons or inclusive charged hadrons. In terms of the $Q_2$ vectors, as defined in Eq.~(\ref{eq:Q2}), $v_2$ values for the particles of interest are found using
\begin{equation} \label{eq:1}
	v_{2,\text{POI}} \{\text{SP}\}=\frac{\langle \QPOI Q^*_{2\mathrm{A}}\rangle}{\sqrt{\frac{\langle Q_{2\mathrm{A}} Q^*_{2\mathrm{B}} \rangle \langle Q_{2\mathrm{A}} Q^*_{2\mathrm{C}} \rangle}{\langle Q_{2\mathrm{B}} Q^*_{2\mathrm{C}} \rangle}}},
\end{equation}
where the subscripts A, B and C refer to different $\eta$-dependent subevents. The $Q^*_2$ vectors are the complex conjugates of the respective $Q_2$ vectors. The \QPOI term is the $Q_2$ vector constructed with the particles of interest. The $Q_{2\mathrm{A}}$ and $Q_{2\mathrm{B}}$ vectors are based on the HF towers, and the $Q_{2\mathrm{C}}$ vector is determined using the information from the tracker, with $\abseta<0.75$. To avoid autocorrelations and ensure an $\eta$ gap of at least 3 units, the scalar product uses $A = \text{HF- (HF+)}$, covering the $\eta$ range $-5 < \eta^\text{HF-} < -3$ ($3 < \eta^\text{HF+} < 5$), and $B = \text{HF+ (HF-)}$, when the charged-hadron track $\eta$ or the reconstructed \PDz meson $y$ is positive (negative). The weight $w_j$ (in Eq.~(\ref{eq:Q2})) is a dimensionless quantity and corresponds to the \ET deposited in individual HF towers for $Q_{2\mathrm{A}}$ and $Q_{2\mathrm{B}}$ and the \pt of the tracks for the tracker ($Q_{2\mathrm{C}}$). A unit weight is assumed for \QPOI. The averages $\langle \rangle$ are calculated considering all events within a given centrality class above a given threshold of \pt for the tracker and \ET for the HF towers. The inclusive charged-hadron $v_2$ is measured based on all charged particle tracks with $1<\pt<3\GeV$ and $\abseta<1$. The \PDz meson $v_2$ is measured in the \pt ranges of 2--4, 4--6, 6--10, and 10--30\GeV with $\absy<1$. Flattening and recentering corrections are applied to the $Q_{2\mathrm{A}}$, $Q_{2\mathrm{B}}$ and $Q_{2\mathrm{C}}$ vectors to account for detector acceptance effects~\cite{flatten1,flatten2}.

\subsection{\texorpdfstring{Prompt \PDz meson $v_2$ measurement}{Prompt D0 meson v2 measurement}}
\label{sec:prompt_D0_v2}
To extract the \PDz meson $v_2$ value for a given $q_2$ bin, it is necessary to correct for the $v_2$ value associated with the underlying background. This is done by applying a modified SP analysis. The individual \PDz candidates are first binned according to their contribution to the SP average, \pPDz, with
\begin{equation} 
	\pPDz = \frac{Q_{2,\PDz} Q^*_{2\mathrm{A}}}{\sqrt{\frac{\langle Q_{2\mathrm{A}} Q^*_{2\mathrm{B}} \rangle \langle Q_{2\mathrm{A}} Q^*_{2\mathrm{C}} \rangle}{\langle Q_{2\mathrm{B}} Q^*_{2\mathrm{C}} \rangle}}}.
\end{equation}
A separate mass spectrum is generated in each of these bins and an invariant mass fit is performed to obtain the associated \PDz meson yield. The \PDz candidates are split into 62 \pPDz bins, with values ranging from $-15$ to $15$, and the bin edges are determined such that the significance of the \PDz signal, as defined in Section~\ref{sec:sre}, is larger than three in each bin. The edges of the \pPDz distribution is chosen such that the number of candidates beyond these limits is insignificant. The inclusive \PDz meson $v_2$ value (\vPDz) is then obtained by taking a yield weighted average of the midpoint of each $p_i(\PDz)$ bin, $p_{i,\text{mid}}$, with 
\begin{equation} \label{eq:v2}
    \vPDz = \frac{\sum p_{i,\text{mid}} Y_i}{\sum Y_i},
\end{equation}
where $Y_i$ is the yield in the $i$th bin. The statistical uncertainty associated with \vPDz is derived by propagating the uncertainties of $Y_i$ in Eq.~(\ref{eq:v2}) as in the Eq.~(\ref{eq:v2_err}),
\begin{equation} \label{eq:v2_err}
    \Delta \vPDz = \frac{\sqrt{\sum \left(p_{i,\text{mid}}-\vPDz\right)^2 (\Delta Y_{i})^{2}}}{\sum Y_i},    
\end{equation}
where $\Delta Y_i$ is the uncertainty in the yield in the $i$th bin.

The \PDz signal in data includes both prompt and nonprompt \PDz contributions. The DCA to the primary vertex of \PDz candidates is used to discriminate between the prompt and nonprompt components. Invariant mass spectra are produced for \PDz mesons corresponding to a centrality class binned in DCA ranges. After obtaining the signal yields based on fits as discussed in Section~\ref{sec:sre}, these yields are used to create a DCA distribution. Templates of prompt and nonprompt \PDz distributions from MC signal simulations are used to fit the yields, with the relative yield parameters as the fit parameters. The fits are conducted using various DCA resolutions applied to the MC template to address discrepancies between the data and the MC simulations. The fit result with the minimum $\chi^2$ is selected as the nominal value for the prompt \PDz fraction. A selection of $\text{DCA}<0.0085\unit{cm}$ is applied, and the resulting sample is estimated to have a 95\% prompt \PDz purity. The systematic uncertainty due to the contribution from nonprompt \PDz is discussed in Section~\ref{sec:syst}.

\subsection{Event shape engineering}\label{sec:ESE}
An event-shape engineering technique is used to classify events based on the initial-state geometry of the collision. The HF towers in the range $3<\abseta<5$ are used for calculating the asymmetry parameter $q_2$. This results in a large $\eta$ gap with the tracks used for \PDz candidate reconstruction, which are restricted to the range $\abseta<1.2$, thus suppressing nonflow correlations. Using both the HF+ and HF- detectors, $q_2$ is defined as $(\abs{Q_2}_\text{HF+}+\abs{Q_2}_\text{HF-})/(\sum_{j=1}^M w_j)$, where $\sum_{j=1}^M w_j$ is the sum of weights from both detectors. The $q_2$ distribution is dependent on centrality and can vary significantly over centrality classes covering 10\% ranges. This introduces a centrality bias where, \eg, events with smaller $q_2$ values are dominated by more central events. To reduce this centrality bias when selecting events based on their $q_2$ values, the events are divided into 1\% centrality bins and, subsequently, each of these bins is further divided into ten separate $q_2$ bins. Since the analysis cannot be performed in 1\% centrality classes because of the limited size of the sample, within a larger 10\% centrality range, $q_2$ bins are combined into a single analysis bin, such that, \eg, all events with the lowest 10\% of $q_2$ values in each of the centrality bins 0--1 to 9--10\% are combined into the lowest 10\% $q_2$ events of the centrality class 0--10\%. A similar approach is followed in an analysis by the ALICE Collaboration~\cite{ALICE_ESE1}.

To study the response of prompt \PDz meson flow to event-shape variations, the \PDz meson $v_2$ values are compared with the low-\pt charged particle $v_2$ values in each $q_2$ interval. The charged particle $v_2$ is used as a data-driven proxy for the event-by-event geometry. To quantify the linearity of the correlation plots, the Pearson correlation coefficient ($r$) is calculated for all five centrality classes studied and for different \pt ranges. The Pearson coefficient is given by
\begin{equation}\label{eq:pearson}
    r = \frac{\sum (x_i-\langle x_w\rangle)(y_i-\langle y_w\rangle)/\sigma_i^2}{\sqrt{\sum (x_i-\langle x_w\rangle)^2/\sigma_i^2} \sqrt{\sum (y_i-\langle y_w\rangle)^2/\sigma_i^2}},
\end{equation} 
where the sum is over all points in the scatter plot, $\sigma_i = \sqrt{\smash[b]{(\Delta x_i)^2 +(\Delta y_i)^2}}$, the $x_i$ and $y_i$ are the charged particle and \PDz meson $v_2$ values for each $q_2$ bin, respectively, the $\Delta x_i$ and $\Delta y_i$ are the statistical uncertainties associated with $x_i$ and $y_i$ respectively, and the $\langle x_w\rangle$ and $\langle y_w\rangle$ are the corresponding weighted-average values over all points. The $\sigma_i$ term takes into account the statistical uncertainty of each point in the scatter plot. The coefficient $r=1(-1)$ for complete linear correlation (anticorrelation), and $r = 0$ in the absence of any linear correlation between two parameters. To estimate the statistical uncertainty in $r$, an MC resampling approach was employed. Each \PDz meson $v_2$ value was varied according to a Gaussian distribution centered on its measured value, with a standard deviation corresponding to its statistical uncertainty. This process was iterated 50000 times to generate a distribution of $r$ values. The final statistical uncertainties were determined using the 68.3\% confidence interval of the resulting distribution.

\section{Systematic uncertainty}
\label{sec:syst} 
The sources of systematic uncertainty include the fit to the signal mass distribution, the fit to the background mass distribution, the sorting of events into $q_2$ ranges, the centrality determination, the nonprompt \PDz meson contamination, and the \PDz meson acceptance and efficiency corrections. The systematic uncertainties associated with the extraction of the \PDz signal is obtained by changing the signal probability density function (pdf) from a double Gaussian to a triple Gaussian with the same mean but different widths, and by changing the function for the combinatorial background from a second- to a third-order polynomial. The systematic uncertainty due to how a $q_2$ range is selected for a given event is found by changing the boundaries of the range associated with each $q_2$ bin up and down by the uncertainty associated with the $Q_2$ flow vectors. The systematic uncertainties associated with the centrality selection is obtained by shifting the centrality boundaries based on the energy deposited in the HF towers up and down. The nonprompt \PDz meson fraction after applying the DCA selection is 5\%, as discussed in Section~\ref{sec:prompt_D0_v2}. Since nonprompt \PDz meson $v_2$ values are smaller than those for prompt \PDz mesons in the studied \pt and centrality region, as shown in Ref.~\cite{HIN-21-003}, the maximum relative uncertainty due to the nonprompt contamination in the $v_2$ value is 5\%. This estimation is based on Eq.~(6) from an earlier CMS publication~\cite{method1}. The effect of the efficiency is studied by changing the efficiency correction values. In an alternative approach, the efficiency is based solely on simulation and does not consider any potential differences between the MC and data \pt spectra. The $v_2$ values for \PDz candidates passing the BDT selection, and for all the \PDz candidates, are calculated from the MC samples, and the difference in $v_2$ is quoted as the systematic uncertainty due to BDT selection. In this analysis, no significant effect resulting from the BDT selection is observed. The summary of the uncertainties is provided in Table~\ref{tab:syst_summary}. Each of these potential sources was varied as described, with the difference in the experimental result based on the varied parameter with that found using the nominal value taken as the associated systematic uncertainty for the slope and intercept. The total uncertainty for each experimental result is taken as the quadratic sum of the individual uncertainties. For $r$, the systematic uncertainty is calculated using a similar approach to that of the statistical uncertainty, where each individual \PDz meson $v_2$ coefficient is varied according to a Gaussian distribution centered on its measured value, with a standard deviation corresponding to its total systematic uncertainty. The final systematic uncertainty was determined using the 68.3\% confidence interval of the resulting distribution.

\begin{table}[!ht]
    \centering
    \topcaption{Summary of systematic uncertainties in the absolute differences between the nominal and the alternative analyses for the \PDz meson $v_2$ values, slope, and intercept of the scatter plots.}
        \begin{tabular}{l c c c }
            & \\ \hline
            Source & 			\PDz meson $v_2$ & 	Slope & 	Intercept\\ \hline
            Sig. mass pdf variation & 	$<$0.034 & 	$<$0.37     & 	$<$0.070 \\
            Bkg. mass pdf variation & 	$<$0.038 & 	$<$0.34     & 	$<$0.086 \\
            $q_2$ selection         &	$<$0.031 & 	$<$0.32     & 	$<$0.026 \\ 
            Centrality selection    &	$<$0.048 & 	0.01--0.54  & 	$<$0.094 \\ 
            Nonprompt               &	$<$0.010 & 	0.02--0.81  & 	$<$0.0033 \\
            $A\,\epsilon$           &	$<$0.011 & 	$<$0.18     & 	$<$0.023\\
            Total                   &   0.004--0.049 & 0.06--0.54 & 0.01--0.12 \\\hline
        \end{tabular}
    \label{tab:syst_summary}
\end{table}

\section{Results}
\label{sec:results}
\begin{figure*}[ht!]
    \centering
    \includegraphics[width=0.98\textwidth]{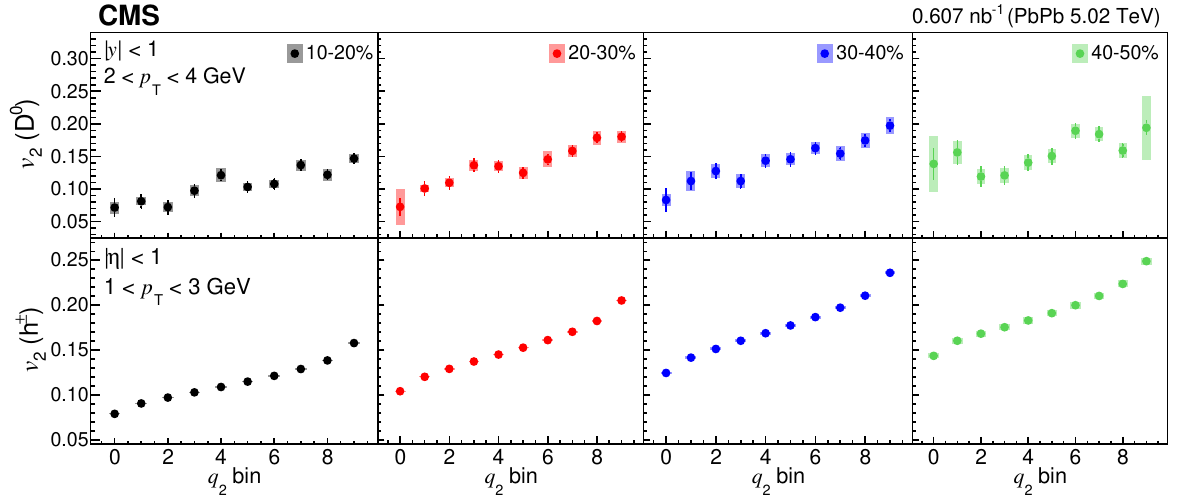}
    \caption{Dependence of the \PDz meson (upper row) and the charged particle (lower row) $v_2$ values on $q_2$. The \PDz mesons have $2<\pt<4\GeV$ and $\absy<1$. The inclusive charged particles have $1<\pt<3\GeV$ and $\abseta<1$. The plots are shown for events in the centrality ranges of 10--20, 20--30, 30--40 and 40--50\%. The vertical lines represent the statistical uncertainties and the boxes represent the systematic uncertainties.}
    \label{fig:ch_v2_vs_q2}
\end{figure*}

\begin{figure*}[ht!]
    \centering
    \includegraphics[width=0.4\textwidth]{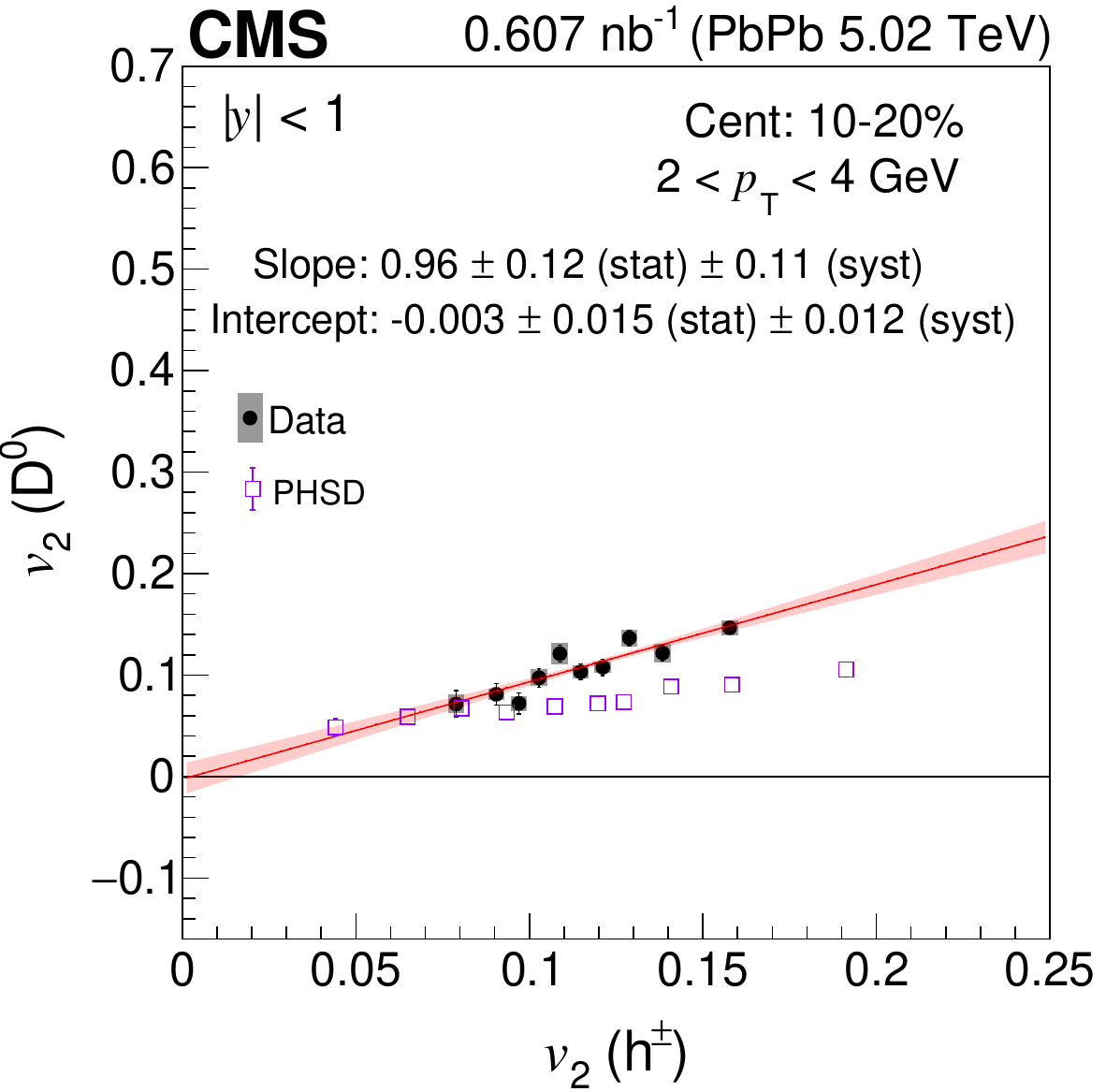}
    \includegraphics[width=0.4\textwidth]{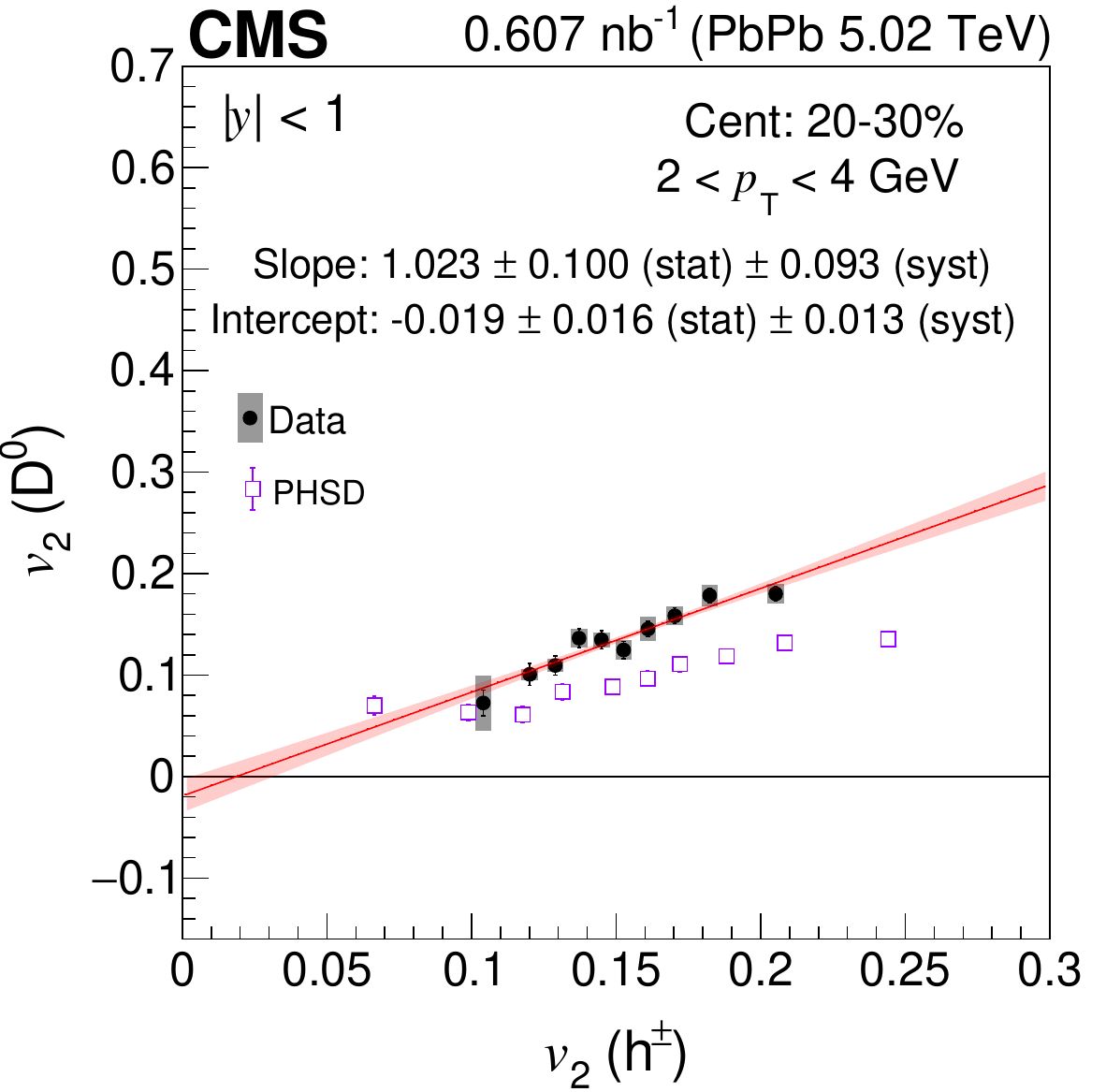}
    \includegraphics[width=0.4\textwidth]{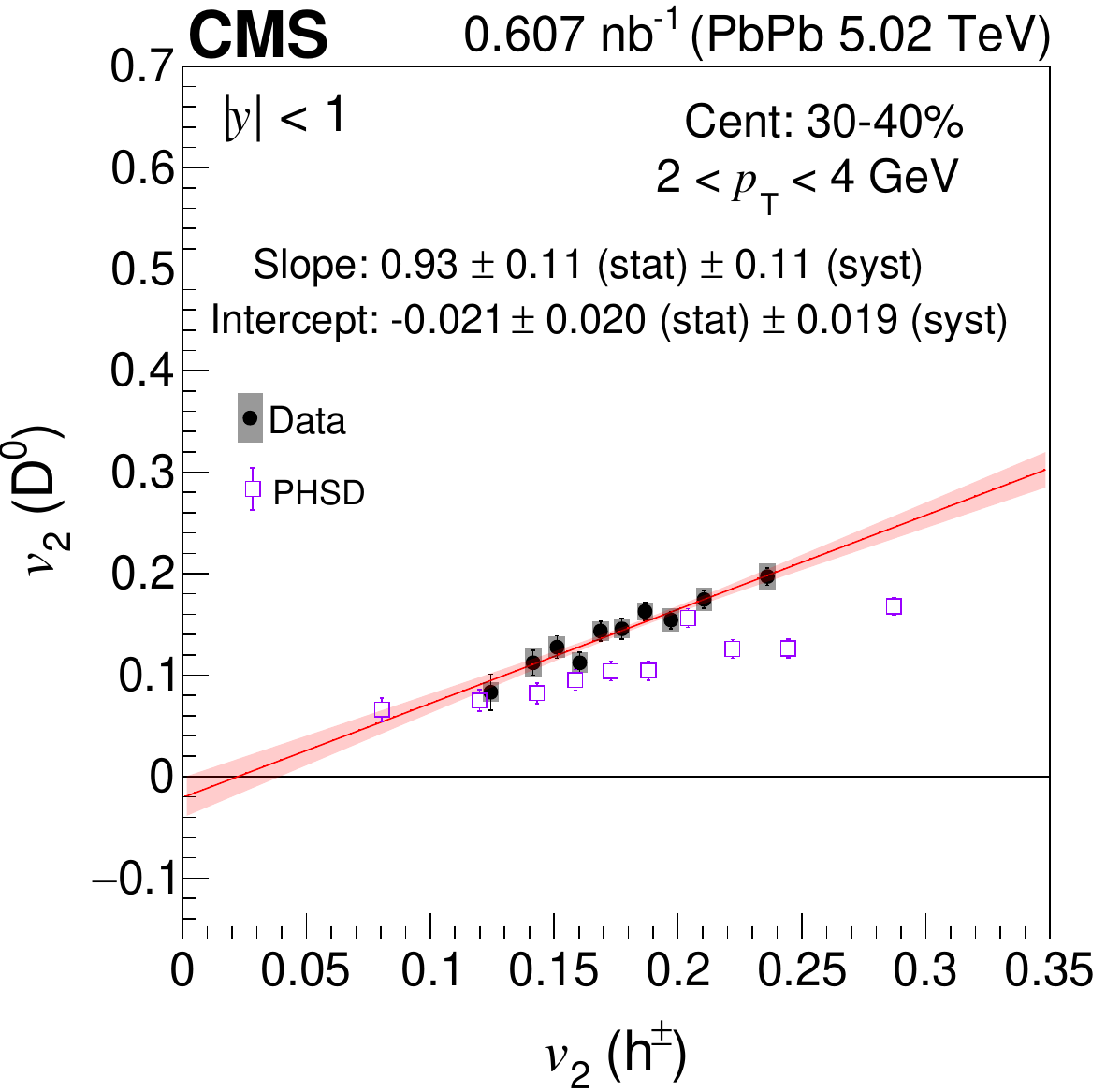}
    \includegraphics[width=0.4\textwidth]{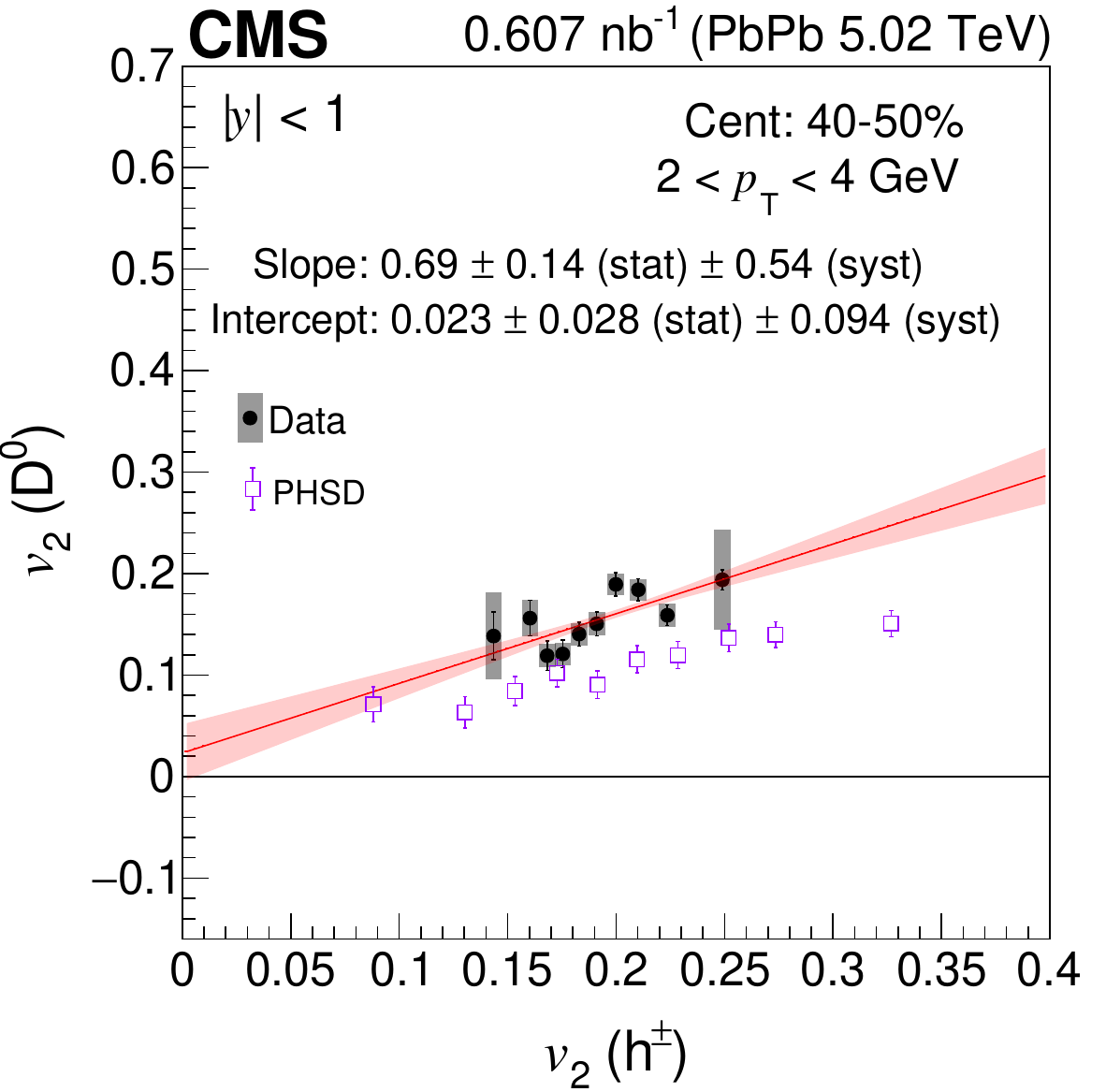}
    \caption{Scatter plots of the \PDz meson $v_2$ \vs charged particle $v_2$ values measured in common $q_2$ intervals. The \PDz mesons have $2<\pt<4\GeV$ and $\absy<1$. The inclusive charged particles have $1<\pt<3\GeV$ and $\abseta<1$. The vertical lines represent the statistical uncertainties, and the boxes represent the systematic uncertainties associated with the \PDz meson $v_2$. The errors associated with the charged particle $v_2$ are smaller than the markers. The centrality classes 10--20\% (upper left), 20--30\% (upper right), 30--40\% (lower left), and 40--50\% (lower right) are shown. The red lines are linear fits to the experimental results based on the statistical uncertainties of the data, with the shaded red band corresponding to a one standard deviation uncertainty. The PHSD model~\cite{PHSD1,PHSD2} predictions are also shown in the plots.}
    \label{fig:corr_1}
\end{figure*}

Figure~\ref{fig:ch_v2_vs_q2} shows the $q_2$ dependence of the $v_2$ values for \PDz mesons (upper row) with $2<\pt<4\GeV$ and $y<1$ and for inclusive charged particles (lower row) with $1<\pt<3\GeV$ and $\abseta<1$. The results are shown for the centrality classes of 10--20, 20--30, 30--40 and 40--50\%. A positive, approximately linear dependence is observed, as expected if both observables are correlated with the event-by-event initial-state eccentricity. Figure~\ref{fig:corr_1} shows the dependence between the \PDz meson $v_2$ values and the corresponding charged particle $v_2$ values with the same $q_2$ percentile classification for selected \PDz \pt ranges and event centrality classes. The observed linear dependence between the \PDz meson and charged particle $v_2$ values indicates that the initial-shape eccentricity is the dominant contributor to the \PDz elliptic flow. Deviations from this linear trend could arise if other physical mechanisms contribute to the \PDz elliptic flow. For instance, a change in the viscosity of the QGP medium can lead to deviations from linearity for more central events~\cite{PhysRevC.95.044901}. A deviation from linear correlation at high \PDz meson \pt values could result from the path-length dependent energy loss of charm quarks in the QGP medium~\cite{PhysRevC.95.044901,PhysRevC.96.064903}. Hints of such deviations from linearity are shown in Appendix~\ref{additional_corr}, where Fig.~\ref{fig:corr_cen_0_10} shows results for the 0--10\% centrality range in several \pt ranges and Fig.~\ref{fig:corr_pT_10_30} shows the results for different centrality ranges with $10<\pt<30\GeV$. At high \pt, the linear correlation between the \PDz meson and inclusive charged particle $v_2$ values becomes weaker, which may be an indication of the path-length dependent energy loss of charm quarks in the QGP medium. 

Figure~\ref{fig:corr_1} also shows the prediction of the parton-hadron-string-dynamics (PHSD) model~\cite{PHSD1,PHSD2}. It is found that the degree of charm-quark thermalization is underestimated in the model, as observed from the lower slope of the scatter plot for \PDz \pt values in the range 2--4\GeV. In this model, the \PDz mesons are produced through initial binary collisions by using the \PYTHIA event generator, accounting for the $(\text{anti-})$~shadowing incorporated in the EPS09 package~\cite{EPS09}.

\begin{figure*}[!htb]
    \centering
    \includegraphics[width=0.8\textwidth]{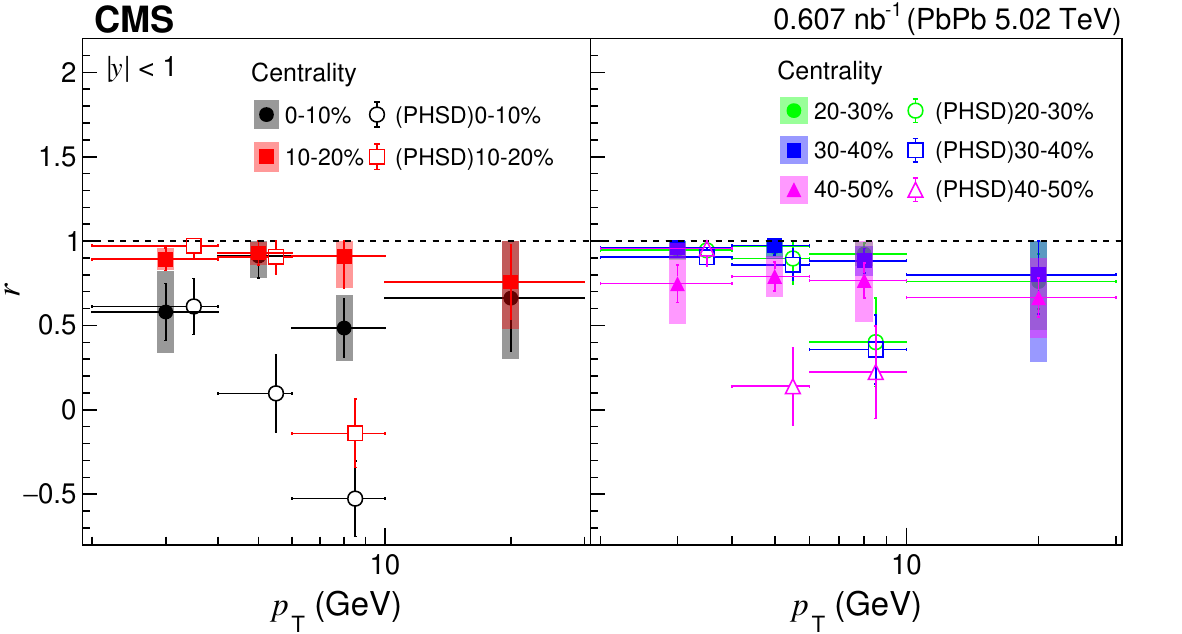} 
    \caption{The Pearson correlation coefficients ($r$) obtained from the scatter plots of the prompt \PDz meson $v_2$ \vs charged particle $v_2$ values for the indicated \pt and centrality classes. The vertical lines correspond to the statistical uncertainties and the vertical bands correspond to the systematic uncertainties added in quadrature. The open symbols show the PHSD predictions in the \PDz meson \pt range  2--10\GeV, with the symbols shifted along the \pt axis for better visibility.}
    \label{fig:pearson}    
\end{figure*}

To extract the linearity of the correlations, we calculate the Pearson correlation coefficients ($r$) as defined in Eq.~(\ref{eq:pearson}). Figure~\ref{fig:pearson} shows the coefficients as function of \pt for different centrality classes. Other than for the most central events, it is observed that the experimental correlation for low-\pt values ($\pt<10\GeV$) is consistent with unity. For the \pt range of 10--30\GeV, the coefficients tend to be less than unity, although with larger uncertainties. The model predictions are consistent with data for $\pt<4\GeV$, whereas the predictions deviate from 1 for $\pt>4\GeV$ for centralities 0--10 and 40--50\%.

\begin{figure*}[!htb]
    \centering
    \includegraphics[width=0.8\textwidth]{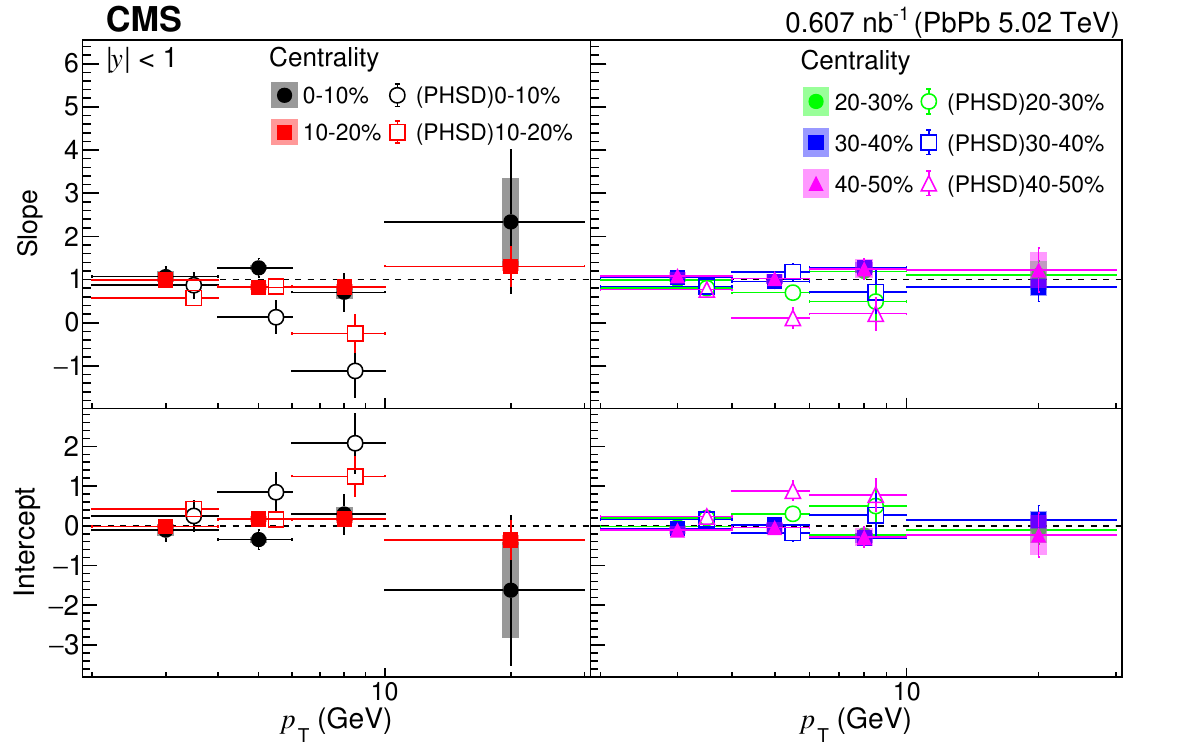} 
    \caption{Slopes (upper row) and intercepts (lower row) obtained from the fits of the prompt \PDz meson \vs charged particle $v_2$ values for each \pt and centrality class. The \PDz meson and charged particle $v_2$ values in each $q_2$ interval are normalized by the corresponding $q_2$-inclusive $v_2$ value in the same \pt and centrality class. The vertical lines correspond to the statistical uncertainties and the vertical bands correspond to the systematic uncertainties added in quadrature. The open symbols show the PHSD predictions in the \PDz meson \pt range 2--10\GeV, with the symbols shifted along the \pt axis for better visibility.}
    \label{fig:slope_intercept_with_syst}
\end{figure*}

To further quantify the linear correlation, Fig.~\ref{fig:slope_intercept_with_syst} presents the slopes and intercepts of the \PDz meson $v_2$ \vs charged particle $v_2$ as a function of \pt for centrality classes 0--10, 10--20, 20--30, 30--40, and 40--50\% in the \PDz meson \pt range of 2--30\GeV. The prompt \PDz meson and charged particle $v_2$ values in each $q_2$ interval are normalized using the corresponding $v_2$ value for the $q_2$-inclusive events (all events in a given 10\% centrality class). This normalization is performed to account for the dependence of these values on \pt and event centrality. The normalized slopes are consistent with unity for all \pt and centrality classes. The intercepts, representing the \PDz meson $v_2$ value when the charged particle $v_2$ value is zero, are also consistent with zero. These slope and intercept results strongly suggest that the \PDz meson elliptic flow originates primarily from the eccentricity of the initial geometry of the collision system. The predictions from PHSD model are consistent with data at low \pt, and at higher \pt the predictions seem to deviate from data for 0--10\% central events.

\section{Summary}
The elliptic-flow $v_2$ Fourier coefficient is measured for \PDz mesons using an event-shape engineering technique. The data sample consists of lead-lead collisions measured with the CMS detector at the CERN LHC at a center-of-mass energy per nucleon pair of 5.02\TeV and with an integrated luminosity of 0.607\nbinv. The azimuthal asymmetry of energy deposits in a forward-rapidity calorimeter is characterized by a $q_2$ parameter that can be related to the initial collision geometry. A strong linear correlation is found between this parameter and the corresponding $v_2$ values for inclusive charged particles measured in the pseudorapidity ($\eta$) and transverse momentum (\pt) ranges of $\abseta<1$ and $1<\pt<3\GeV$, respectively. The variation between \PDz meson and inclusive charged particle $v_2$ values obtained at common $q_2$ intervals is then explored. A strong linear correlation is found over the centrality range of 10--40\%, as binned in 10\% classes, for \PDz mesons at low \pt (2--10\GeV). A weaker linear correlation is observed for centrality ranges of 0--10 and 40--50\%. The parton-hadron-string-dynamics model is compared to the experimental results and found to underestimate the degree of charm quark thermalization in the quark-gluon plasma at $\pt<4\GeV$. These findings provide compelling evidence for the importance of the initial-state geometry in determining the collective flow of heavy quarks within the quark-gluon plasma created in heavy ion collisions.
\begin{acknowledgments}
We congratulate our colleagues in the CERN accelerator departments for the excellent performance of the LHC and thank the technical and administrative staffs at CERN and at other CMS institutes for their contributions to the success of the CMS effort. In addition, we gratefully acknowledge the computing centers and personnel of the Worldwide LHC Computing Grid and other centers for delivering so effectively the computing infrastructure essential to our analyses. Finally, we acknowledge the enduring support for the construction and operation of the LHC, the CMS detector, and the supporting computing infrastructure provided by the following funding agencies: SC (Armenia), BMFWF and FWF (Austria); FNRS and FWO (Belgium); CNPq, CAPES, FAPERJ, FAPERGS, and FAPESP (Brazil); MES and BNSF (Bulgaria); CERN; CAS, MoST, and NSFC (China); MINCIENCIAS (Colombia); MSES and CSF (Croatia); RIF (Cyprus); SENESCYT (Ecuador); ERC PRG and PSG, TARISTU24-TK10 and MoER TK202 (Estonia); Academy of Finland, MEC, and HIP (Finland); CEA and CNRS/IN2P3 (France); SRNSF (Georgia); BMFTR, DFG, and HGF (Germany); GSRI (Greece); MATE and NKFIH (Hungary); DAE and DST (India); IPM (Iran); SFI (Ireland); INFN (Italy); MSIT and NRF (Republic of Korea); MES (Latvia); LMTLT (Lithuania); MOE and UM (Malaysia); BUAP, CINVESTAV, CONACYT, LNS, SEP, and UASLP-FAI (Mexico); MOS (Montenegro); MBIE (New Zealand); PAEC (Pakistan); MSHE, NSC, and NAWA (Poland); FCT (Portugal); MESTD (Serbia); MICIU/AEI and PCTI (Spain); MOSTR (Sri Lanka); Swiss Funding Agencies (Switzerland); MST (Taipei); MHESI (Thailand); TUBITAK and TENMAK (T\"{u}rkiye); NASU (Ukraine); STFC (United Kingdom); DOE and NSF (USA).

\begin{sloppypar}
\setlength\emergencystretch{\hsize}
\hyphenation{Rachada-pisek} Individuals have received support from the Marie-Curie program and the European Research Council and Horizon 2020 Grant, contract Nos.\ 675440, 724704, 752730, 758316, 765710, 824093, 101115353, 101002207, 101001205, and COST Action CA16108 (European Union); the Leventis Foundation; the Alfred P.\ Sloan Foundation; the Alexander von Humboldt Foundation; the Science Committee, project no. 22rl-037 (Armenia); the Fonds pour la Formation \`a la Recherche dans l'Industrie et dans l'Agriculture (FRIA) and Fonds voor Wetenschappelijk Onderzoek contract No. 1228724N (Belgium); the Beijing Municipal Science \& Technology Commission, No. Z191100007219010, the Fundamental Research Funds for the Central Universities, the Ministry of Science and Technology of China under Grant No. 2023YFA1605804, the Natural Science Foundation of China under Grant No. 12535004, and USTC Research Funds of the Double First-Class Initiative No.\ YD2030002017 (China); the Ministry of Education, Youth and Sports (MEYS) of the Czech Republic; the Shota Rustaveli National Science Foundation (Georgia); the Deutsche Forschungsgemeinschaft (DFG), among others, under Germany's Excellence Strategy -- EXC 2121 ``Quantum Universe" -- 390833306, and under project number 400140256 - GRK2497; the Hellenic Foundation for Research and Innovation (HFRI), Project Number 2288 (Greece); the Hungarian Academy of Sciences, the New National Excellence Program - \'UNKP, the NKFIH research grants K 131991, K 138136, K 143460, K 143477, K 147557, K 146913, K 146914, K 147048, TKP2021-NKTA-64, and 2025-1.1.5-NEMZ\_KI-2025-00004, and MATE KKP and KKPCs Research Excellence and Flagship Research Groups grants (Hungary); the Council of Science and Industrial Research, India; ICSC -- National Research Center for High Performance Computing, Big Data and Quantum Computing, FAIR -- Future Artificial Intelligence Research, and CUP I53D23001070006 (Mission 4 Component 1), funded by the NextGenerationEU program, the Italian Ministry of University and Research (MUR) under Bando PRIN 2022 -- CUP I53C24002390006, PRIN PRIMULA 2022RBYK7T (Italy); the Latvian Council of Science; the Ministry of Science and Higher Education, project no. 2022/WK/14, and the National Science Center, contracts Opus 2021/41/B/ST2/01369, 2021/43/B/ST2/01552, 2023/49/B/ST2/03273, and the NAWA contract BPN/PPO/2021/1/00011 (Poland); the Funda\c{c}\~ao para a Ci\^encia e a Tecnologia (Portugal); the National Priorities Research Program by Qatar National Research Fund; MICIU/AEI/10.13039/501100011033, ERDF/EU, ``European Union NextGenerationEU/PRTR", projects PID2022-142604OB-C21, PID2022-139519OB-C21, PID2023-147706NB-I00, PID2023-148896NB-I00, PID2023-146983NB-I00, PID2023-147115NB-I00, PID2023-148418NB-C41, PID2023-148418NB-C42, PID2023-148418NB-C43, PID2023-148418NB-C44, PID2024-158190NB-C22, RYC2021-033305-I, RYC2024-048719-I, CNS2023-144781, CNS2024-154769 and Plan de Ciencia, Tecnolog{\'i}a e Innovaci{\'o}n de Asturias, Spain; the Chulalongkorn Academic into Its 2nd Century Project Advancement Project, the National Science, Research and Innovation Fund program IND\_FF\_68\_369\_2300\_097, and the Program Management Unit for Human Resources \& Institutional Development, Research and Innovation, grant B39G680009 (Thailand); the Eric \& Wendy Schmidt Fund for Strategic Innovation through the CERN Next Generation Triggers project under grant agreement number SIF-2023-004; the Kavli Foundation; the Nvidia Corporation; the SuperMicro Corporation; the Welch Foundation, contract C-1845; and the Weston Havens Foundation (USA).
\end{sloppypar}
\end{acknowledgments}\section*{Data availability} Release and preservation of data used by the CMS Collaboration as the basis for publications is guided by the  \href{https://doi.org/10.7483/OPENDATA.CMS.1BNU.8V1W}{CMS data preservation, re-use and open access policy}.
\bibliography{auto_generated} 

\appendix
\section{\texorpdfstring{Scatter plots of the \PDz meson $v_2$ \vs charged particle $v_2$}{Scatter plots of the D0 meson v2 vs. charged particle v2}}
\label{additional_corr}
\begin{figure*}[ht!b]
    \centering
    \includegraphics[width=0.4\textwidth]{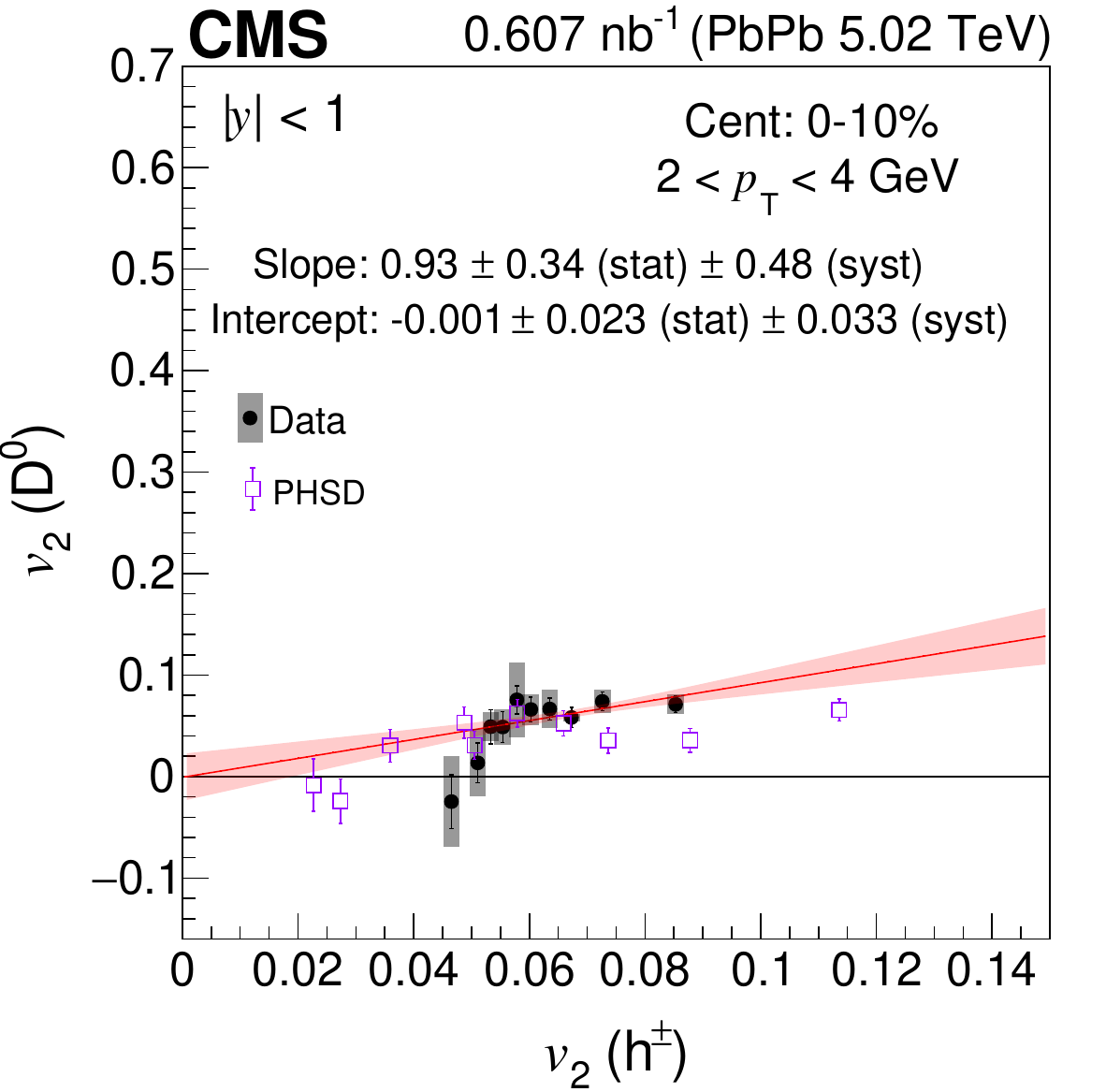}
    \includegraphics[width=0.4\textwidth]{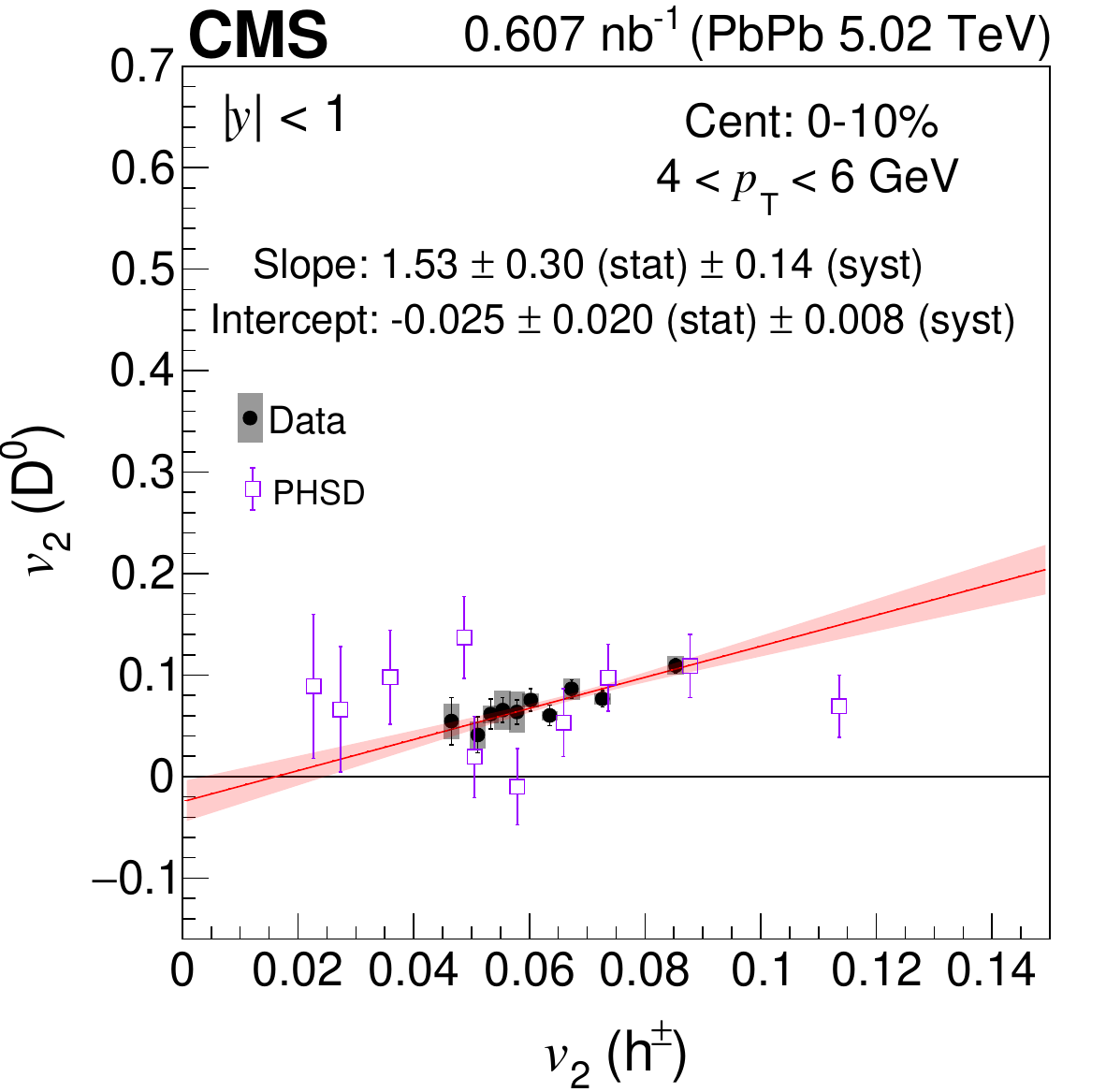}
    \includegraphics[width=0.4\textwidth]{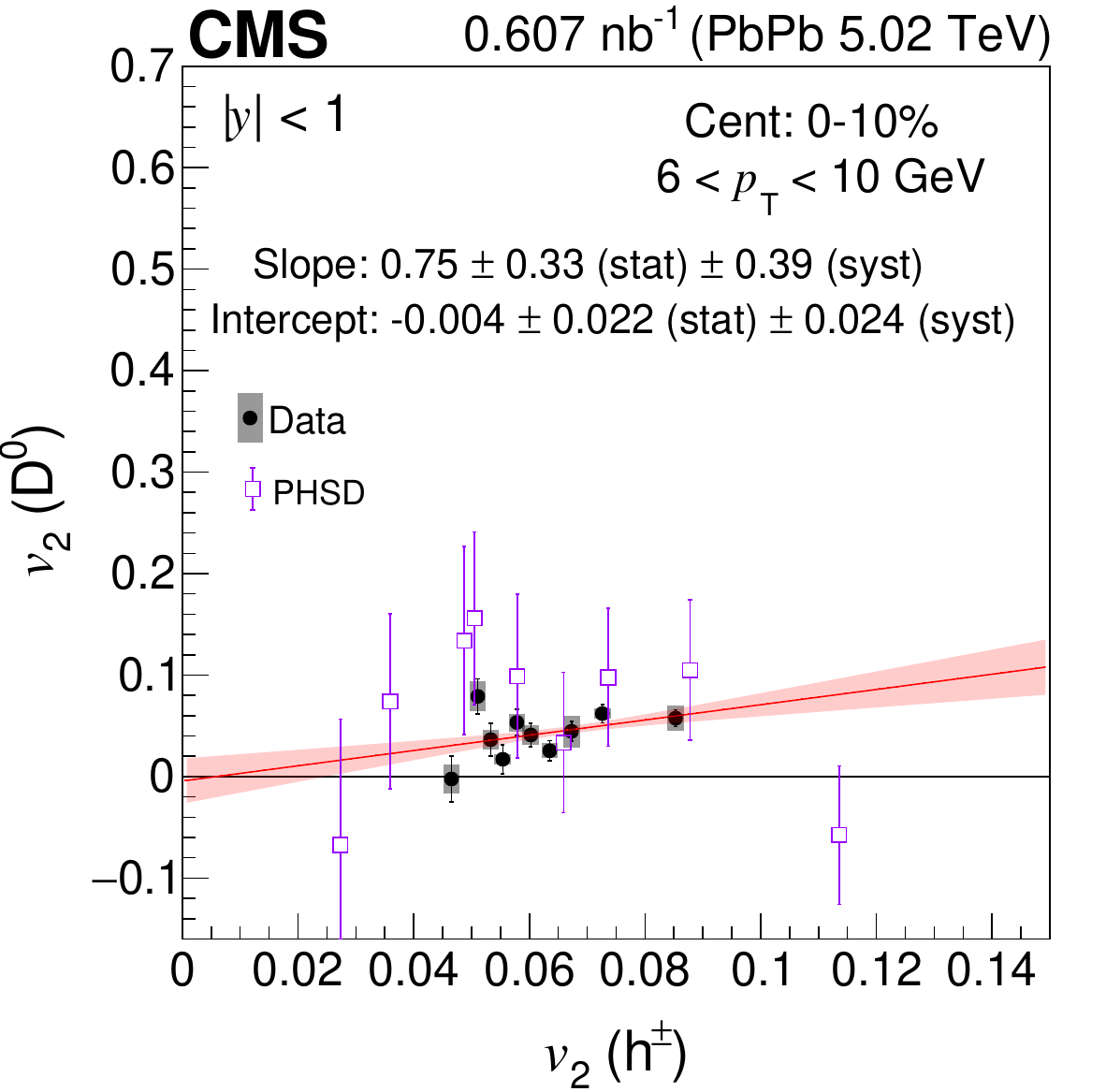}
    \includegraphics[width=0.4\textwidth]{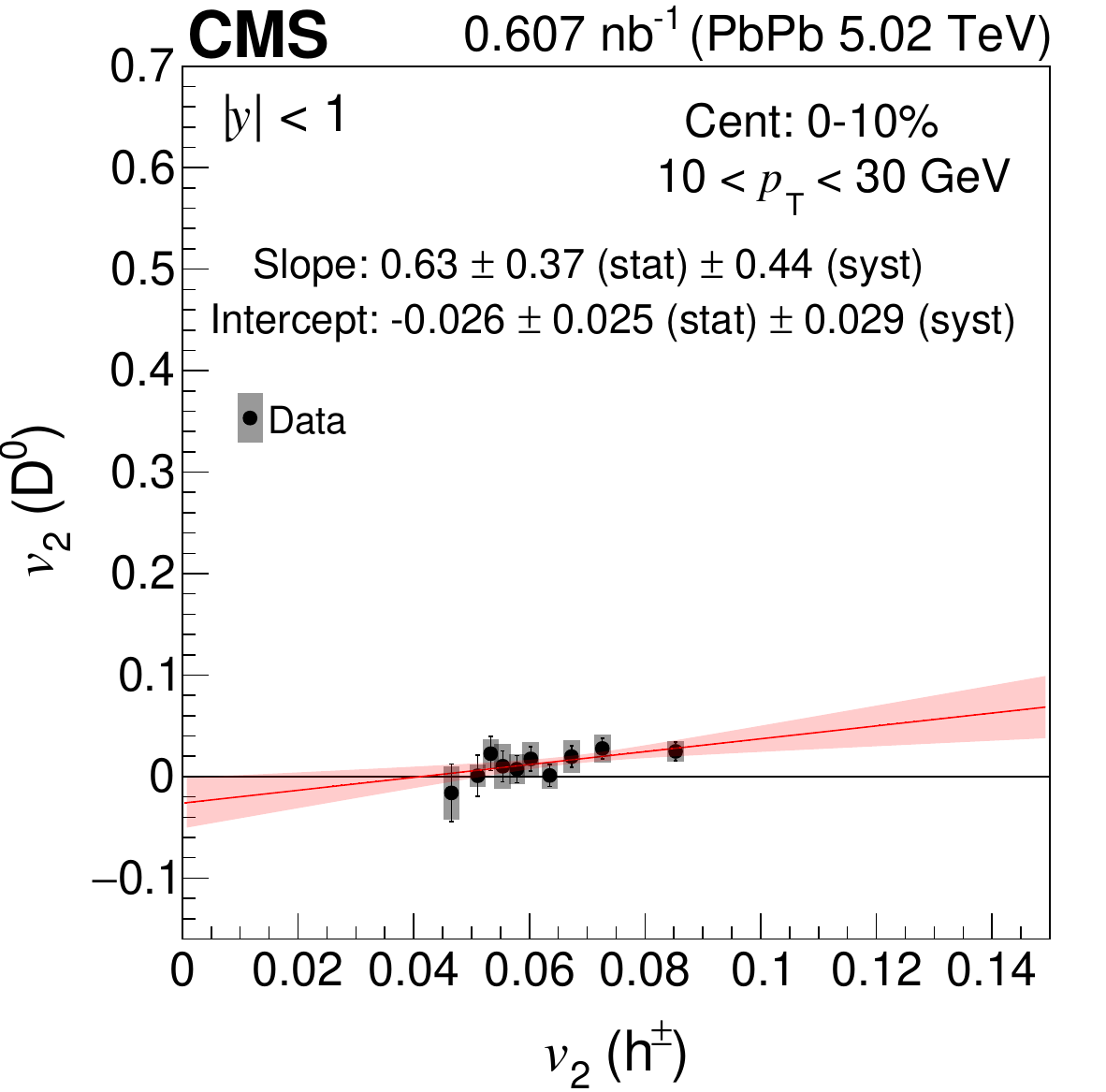}
    \caption{Scatter plots of the \PDz meson $v_2$ \vs charged particle $v_2$ values measured in common $q_2$ intervals. The inclusive charged particles have $1<\pt<3\GeV$ and $\abseta<1$. The vertical lines represent the statistical uncertainties, and the boxes represent the systematic uncertainties associated with the \PDz meson $v_2$ values. The uncertainties associated with the inclusive charged particle $v_2$ values are smaller than the markers. The \PDz meson \pt ranges 2--4\GeV (upper left), 4--6\GeV (upper right), 6--10\GeV (lower left), and 10--30\GeV (lower right) for the 0--10\% centrality class are shown. The red lines are linear fits to the experimental results based on the statistical uncertainties of the data, with the shaded red band corresponding to a one standard deviation uncertainty. The PHSD model~\cite{PHSD1,PHSD2} predictions are also shown in the plots.}
    \label{fig:corr_cen_0_10}
\end{figure*}

\begin{figure*}[ht!b]
    \centering
    \includegraphics[width=0.4\textwidth]{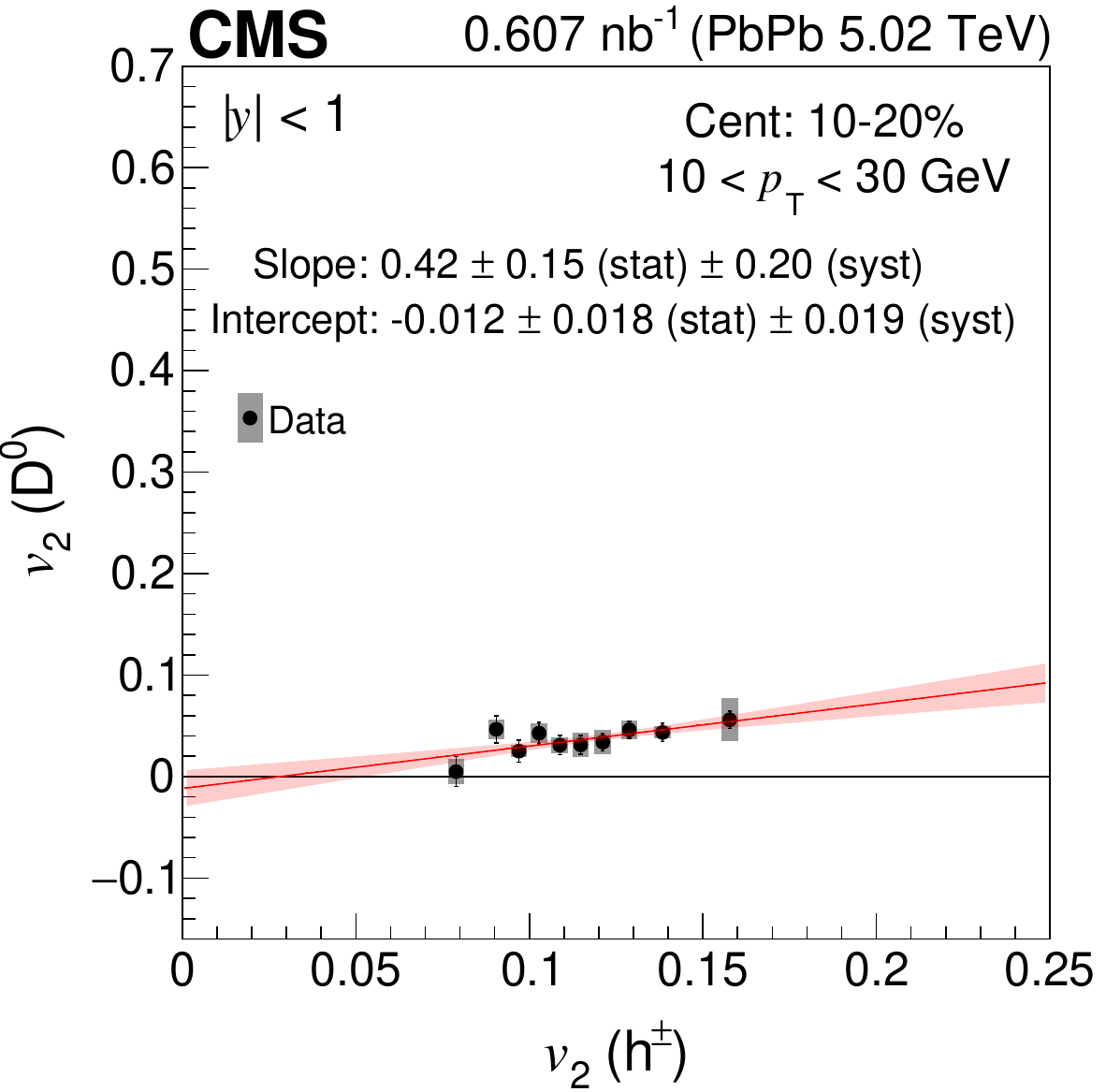}
    \includegraphics[width=0.4\textwidth]{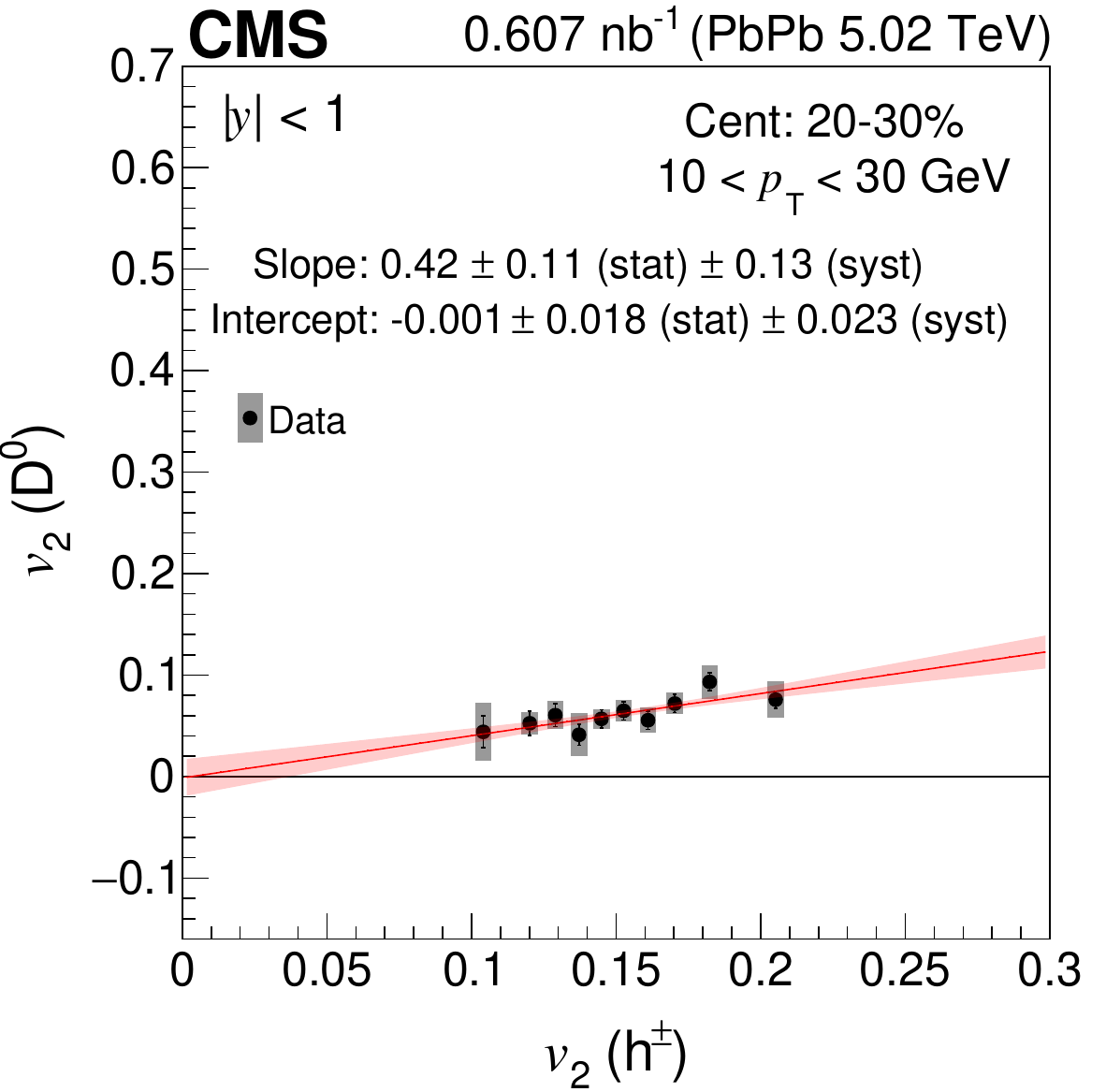}
    \includegraphics[width=0.4\textwidth]{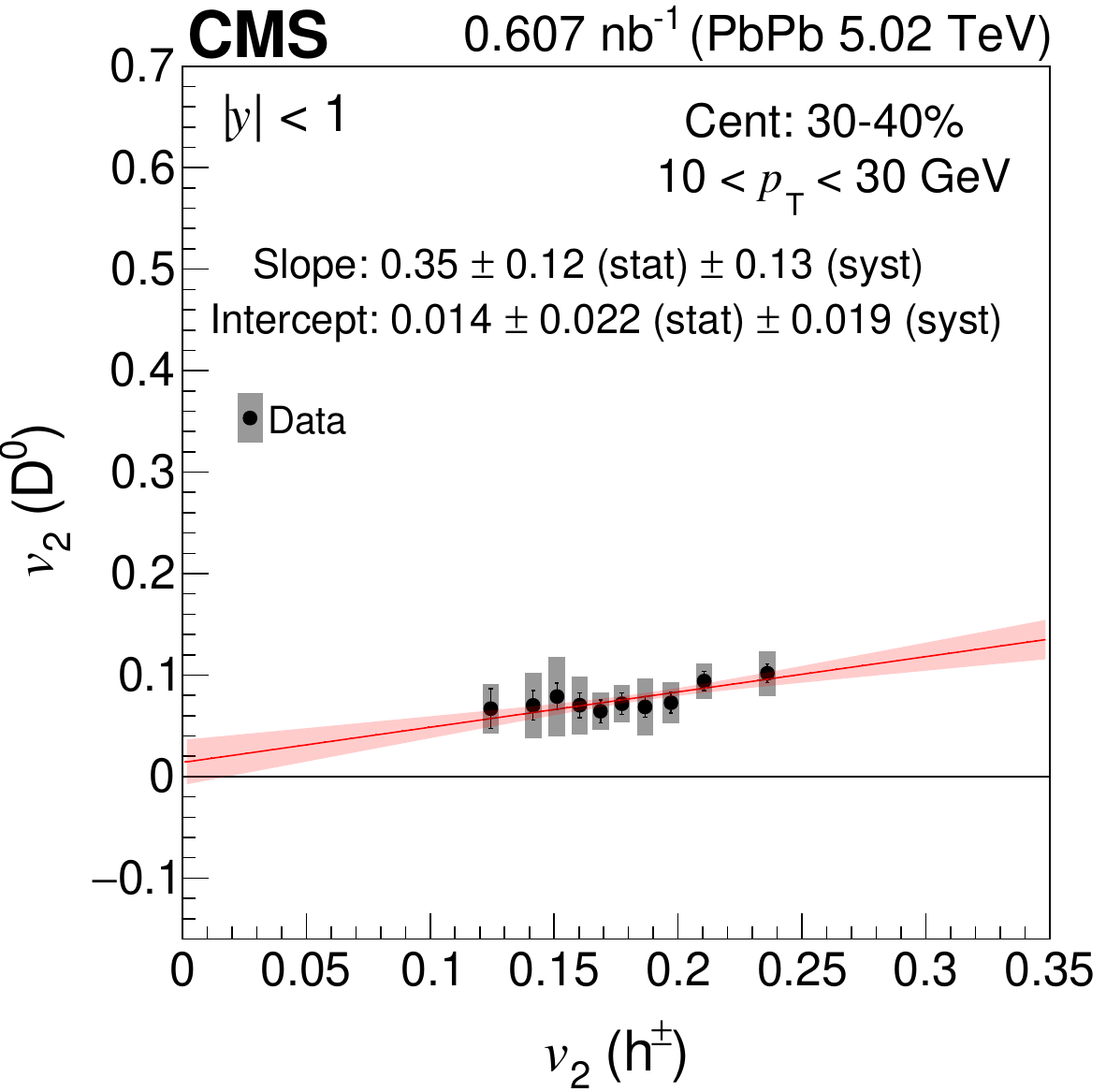}
    \includegraphics[width=0.4\textwidth]{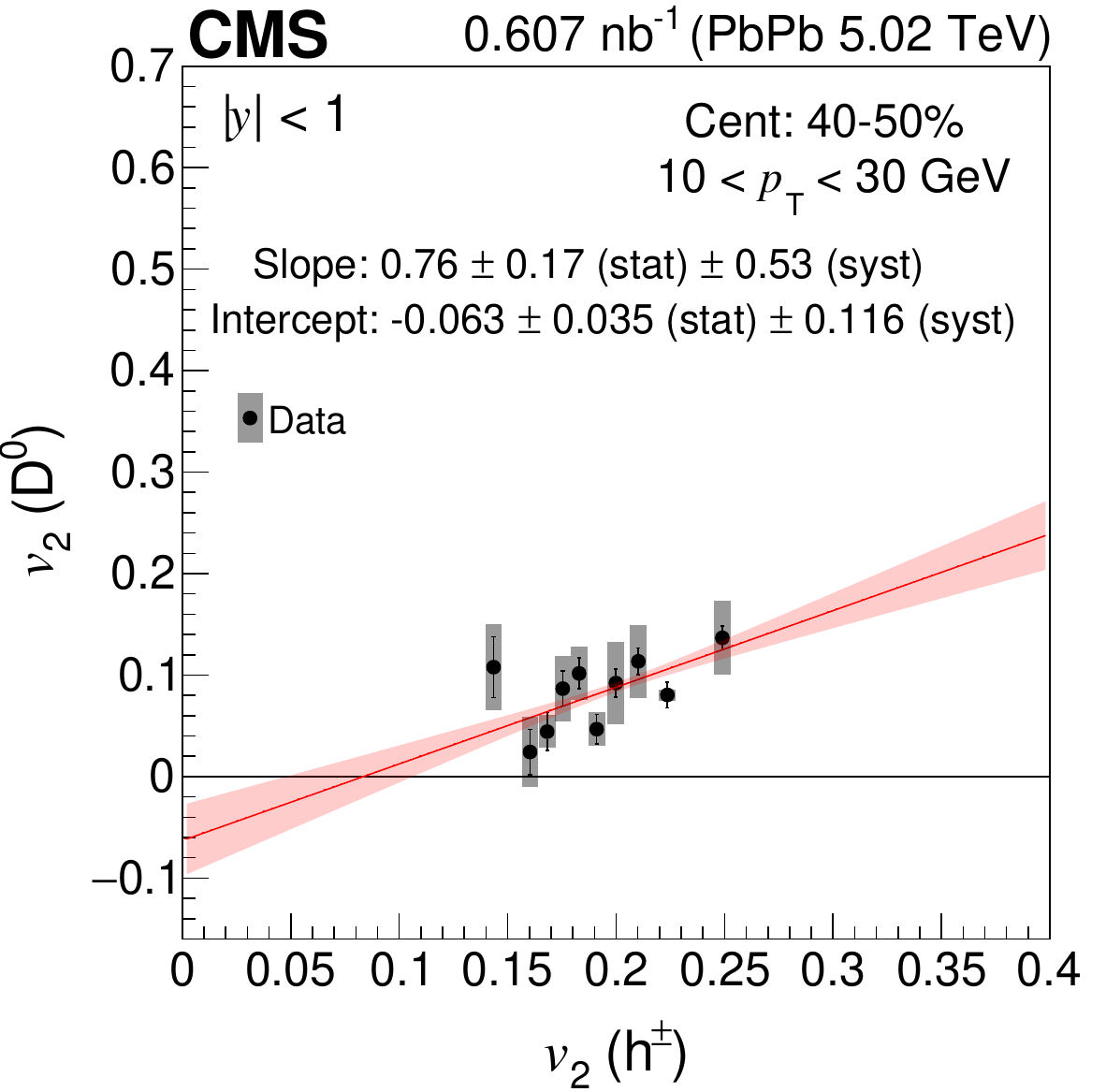}
    \caption{ The same scatter plots as Fig.~\ref{fig:corr_cen_0_10}, but for the \PDz mesons with $10<\pt<30\GeV$ in the 10--20, 20--30, 30--40, and 40--50\% centrality classes. All other explanation for the plots are the same as in Fig.~\ref{fig:corr_cen_0_10}.}
    \label{fig:corr_pT_10_30}
\end{figure*}

The scatter plots of the \PDz meson $v_2$ \vs charged particle $v_2$ values measured in common $q_2$ intervals for the most central events, \ie, 0--10\%, are shown in Fig.~\ref{fig:corr_cen_0_10}. The scatter plots for \PDz meson $\pt>10\GeV$ are shown in Fig.~\ref{fig:corr_pT_10_30}.
\cleardoublepage \section{The CMS Collaboration \label{app:collab}}\begin{sloppypar}\hyphenpenalty=5000\widowpenalty=500\clubpenalty=5000\cmsinstitute{Yerevan Physics Institute, Yerevan, Armenia}
{\tolerance=6000
A.~Gevorgyan\cmsorcid{0000-0003-2751-9489}, A.~Hayrapetyan, V.~Makarenko\cmsorcid{0000-0002-8406-8605}, A.~Tumasyan\cmsAuthorMark{1}\cmsorcid{0009-0000-0684-6742}
\par}
\cmsinstitute{Marietta Blau Institute for Particle Physics, Vienna, Austria}
{\tolerance=6000
W.~Adam\cmsorcid{0000-0001-9099-4341}, L.~Benato\cmsorcid{0000-0001-5135-7489}, T.~Bergauer\cmsorcid{0000-0002-5786-0293}, M.~Dragicevic\cmsorcid{0000-0003-1967-6783}, P.S.~Hussain\cmsorcid{0000-0002-4825-5278}, M.~Jeitler\cmsAuthorMark{2}\cmsorcid{0000-0002-5141-9560}, N.~Krammer\cmsorcid{0000-0002-0548-0985}, A.~Li\cmsorcid{0000-0002-4547-116X}, D.~Liko\cmsorcid{0000-0002-3380-473X}, M.~Matthewman, J.~Schieck\cmsAuthorMark{2}\cmsorcid{0000-0002-1058-8093}, R.~Sch\"{o}fbeck\cmsAuthorMark{2}\cmsorcid{0000-0002-2332-8784}, M.~Shooshtari\cmsorcid{0009-0004-8882-4887}, M.~Sonawane\cmsorcid{0000-0003-0510-7010}, N.~Van~Den~Bossche\cmsorcid{0000-0003-2973-4991}, W.~Waltenberger\cmsorcid{0000-0002-6215-7228}, C.E.~Wulz\cmsAuthorMark{2}\cmsorcid{0000-0001-9226-5812}
% ...
\par}
$^{1}$Also at Yerevan State University, Yerevan, Armenia\\
$^{2}$Also at Technische Universit\"{a}t Wien, Vienna, Austria\\
% ...
\end{sloppypar}
\end{document}